\documentclass{elsart}

\usepackage{natbib}

\usepackage{graphicx}
\usepackage{epsfig}

\usepackage{amssymb}

\begin{document}

\begin{frontmatter}



\title{Spectrum, Intensity and Coherence in Weighted Networks of 
a Financial Market}


\author[label1]{Gergely Tib\'ely,}
\author[label2]{Jukka-Pekka Onnela,}
\author[label2]{Jari Saram\"aki,}
\author[label2]{Kimmo Kaski,}
\author[label1,label2]{J\'anos Kert\'esz}
\address[label1]{Department of Theoretical Physics, Budapest University of Technology and Economics, Budafoki út 8, H-1111 Budapest, Hungary}
\address[label2]{Laboratory Computational Engineering, Helsinki University of Technology, P.O.Box 9203, FIN-02015 HUT}

\begin{abstract}
We construct a correlation matrix based financial network for a set of
New York Stock Exchange (NYSE) traded stocks with stocks corresponding 
to nodes and the links between them added one after the other, 
according to the strength of the correlation between the nodes.  
The eigenvalue spectrum of the correlation matrix reflects the 
structure of the market, which also shows in the cluster structure 
of the emergent network. The stronger and more compact a cluster is, 
the earlier the eigenvalue representing the corresponding business 
sector occurs in the spectrum.  On the other hand, if groups of 
stocks belonging to a given business sector are considered as a 
fully connected subgraph of the final network, their intensity and 
coherence can be monitored as a function of time. This approach 
indicates to what extent the business sector classifications 
are visible in market prices, which in turn enables us to gauge the 
extent of group-behaviour exhibited by stocks belonging to a given 
business sector.

\end{abstract}




\end{frontmatter}

\section{Introduction}
\label{sec:intro}
In the world of business, companies interact with one another,
creating an evolving complex system \cite{santafe}. While the details
of these interactions are not known, as far as price changes are
concerned, they are reflected in the correlations of stock
prices. Correlations are central in investment theory and risk
management, and also serve as inputs to the portfolio optimisation
problem in the classical Markowitz portfolio theory \cite{Mark}.

Complex networks can be seen to provide a general framework for studying 
systems with large numbers of interacting agents~\cite{barabasi}. 
The nodes of the network represent the agents and a link connecting two 
nodes indicates an interaction between them. In this framework, 
interactions have typically been considered binary in nature, 
meaning that two nodes are either connected or not. However, 
in a system with correlated nodes the notion of binary interactions 
implies setting a threshold value for interaction strength, above 
which the link exists and below it does not. This entails a certain 
loss of information as for the properties of the system, which
can be circumvented by assigning weights on the links to reflect 
their interaction strengths. These are naturally identified by the
corresponding elements of the correlation matrix. 

In this paper we study a financial network in which the nodes
correspond to stocks and links to return correlation based
interactions between them. Mantegna~\cite{Man} was the first to
construct such networks and the idea was then followed and 
extended by others~\cite{Van,marsili,caldarelli,jppre,tiziana}.

\section{Network Construction}
\label{sec:method}

We start by considering a price time series for a set of $N$ stocks and 
denote the daily closing price of stock $i$ at time $\tau$ (an actual date) 
by $P_{i}(\tau)$. Here we will analyse $N=116$ NYSE-traded stocks 
from the S\&P 500 index over the period from 1.1.1982 to 31.12.2000, 
consisting the total of 4787 daily closing price quotes for each stock.
As it is common among investors in the financial market, we will work 
in terms of relative returns defined as 
$r_{i}(\tau)=\ln P_{i}(\tau)-\ln P_{i}(\tau-1)$. For the purpose of 
smoothening, we set a moving time window of width $T$, here $T=1000$ 
trading days ($\approx 4$ years, for 250 trading days a year), and 
obtain a return vector $\mathbf{r}_{i}^t$ for stock $i$, where 
the superscript $t$ enumerates the time window under consideration. 
Now the equal time correlation coefficients between assets $i$ and $j$ 
can be written as follows

\begin{equation}
\rho _{ij}^t=\frac{\langle \mathbf{r}_{i}^t \mathbf{r}_{j}^t
  \rangle -\langle \mathbf{r}_{i}^t \rangle \langle \mathbf{r}_{j}^t \rangle }
  {\sqrt{[\langle {\mathbf{r}_{i}^t}^{2} \rangle
      -\langle \mathbf{r}_{i}^t\rangle ^{2}][\langle {\mathbf{r}_{j}^t}^{2} 
       \rangle -\langle \mathbf{r}_{j}^t \rangle
      ^{2}]}}, 
\end{equation}

\noindent where $\left\langle ...\right\rangle $ indicates a time
average over the consecutive trading days included in the return
vectors. These correlation coefficients between $N$ assets form a
symmetric $N\times N$ correlation matrix $\mathbf{C}^t$ with elements 
$\rho _{ij}^t$. The time windows are displaced by $\delta T$, where 
we have used a step size of one trading week, i.e. $\delta T = 5$ 
days. 

We construct the network first by ranking the interaction strengths 
$w_{ij}$ taken as absolute values of the correlation coefficients.  
Due to the fact that $\rho _{ij}^t$ vary between $-1$ and $1$, the 
interaction strengths $w_{ij}^t = |\rho_{ij}^t|$ are limited to the 
$[0,1]$ interval.  Then the network is constructed such that the links 
are added one after the other, starting from the strongest one according 
to the ranking. The emergent network is characterized by a parameter
$p$, namely the ratio of the created links to the number of all
possible links, $N(N-1)/2$. In the end of the procedure when $p=1$, 
we have a fully connected weighted network. In \cite{epjb} we have 
reported this approach and found clear evidence of strong intra-business 
sector clustering for low values of $p$, where we followed the Forbes 
business sector labelling of stocks into 12 categories, such as Energy 
and Utilities \cite{forbes}.

\section{Spectral Properties}
\label{sec:spectrum}

The spectra of financial correlation matrices have been studied in detail, 
producing interesting results \cite{rmt}. The eigenvalues can be 
classified as follows: i) There is a quasi-continuum of small 
eigenvalues which can be well described by the random matrix theory 
corresponding to noise, and the majority of them fall into this category. 
ii) The largest eigenvalue is far from the rest and it corresponds 
to the global behaviour of the market. iii) The discrete spectrum of 
intermediate eigenvalues carries important information about the 
correlations that can be related to market taxonomy. As an example of 
applications, the eigenvalue spectrum can be used to denoise the 
correlation matrix \cite{pafka}. The eigenvalue spectrum also
reflects the business sector structure of the network. Therefore, it is 
natural to ask the question: How do the typical eigenvalues emerge 
as a function of the ratio of the links present or occupation 
$p$? 

Here we calculate the eigenvalue spectrum of the matrix 
$w_{ij}-\delta_{ij}$ for different values of $p$, where $\delta_{ij}$ 
is the Kronecker delta function. Starting from the strongest links 
first the most correlated parts emerge in the network. They form 
separated clusters with high clustering coefficients, thus the emerging 
structure is far from random. The eigenvalue spectrum reflects this 
property. Already for very small values of $p$ the largest eigenvalue 
separates from the rest and the components are quite uniformly distributed 
among the stocks already included in the network, indicating the dominance 
of the global market behaviour. To give the eigenvalues some physical 
meaning it is convenient to plot the values of each of the 116 components 
of the eigenvector corresponding to the chosen eigenvalue.  

\begin{figure}[!h]
\centering
\includegraphics[width=160pt]{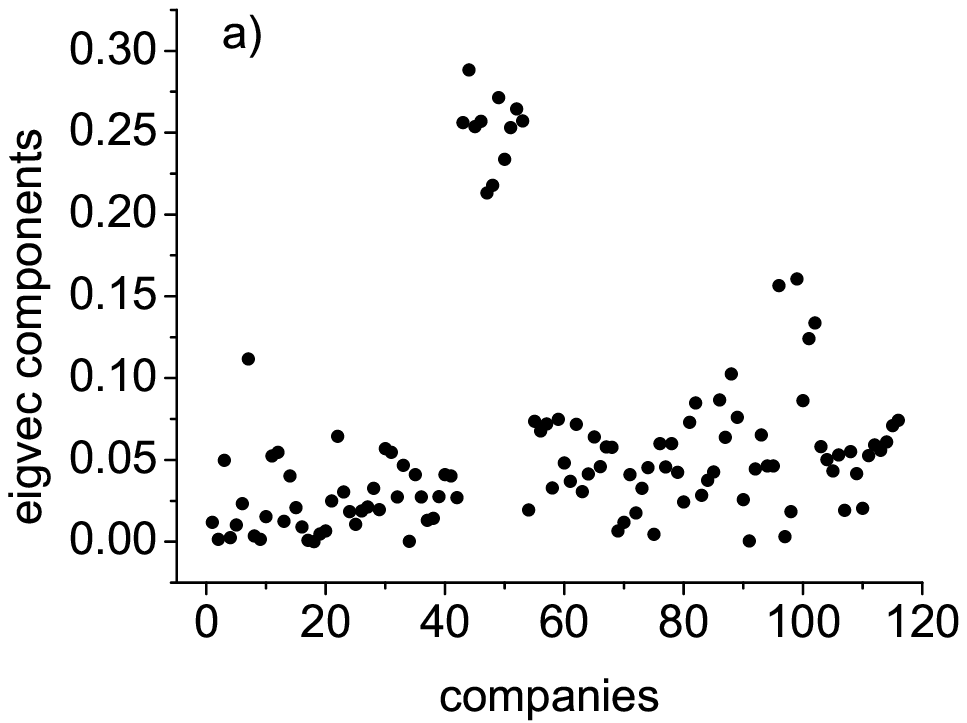}
\includegraphics[width=160pt]{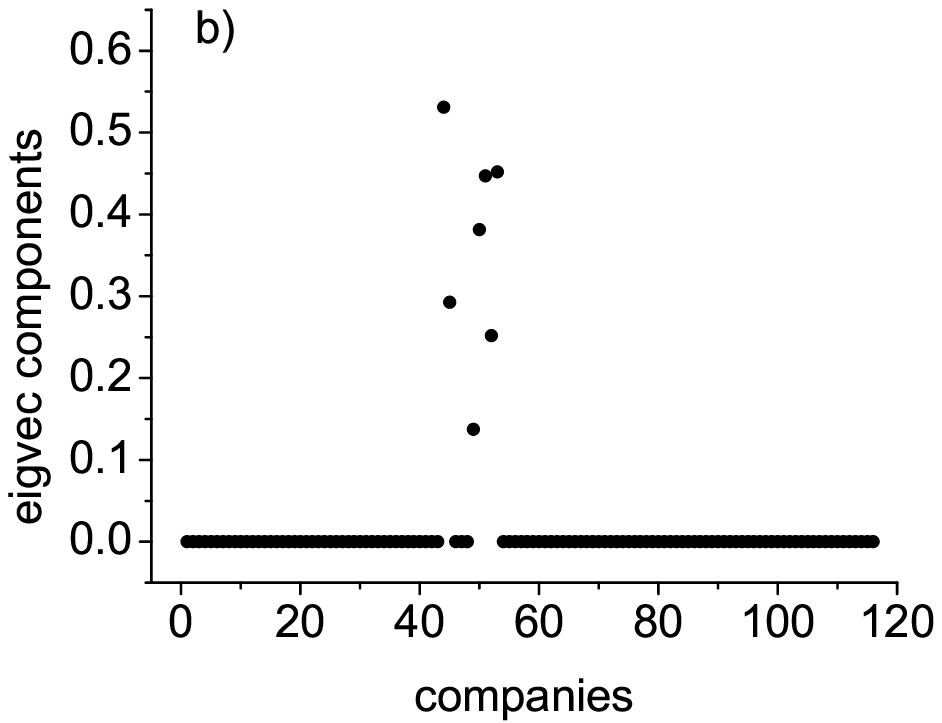}
\includegraphics[width=160pt]{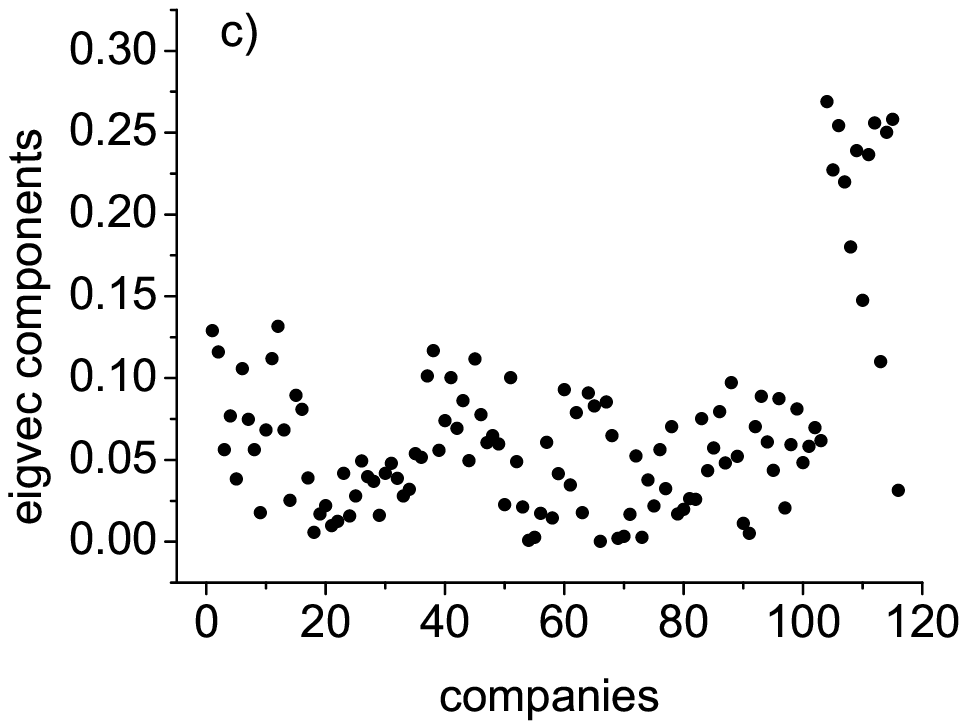}
\includegraphics[width=160pt]{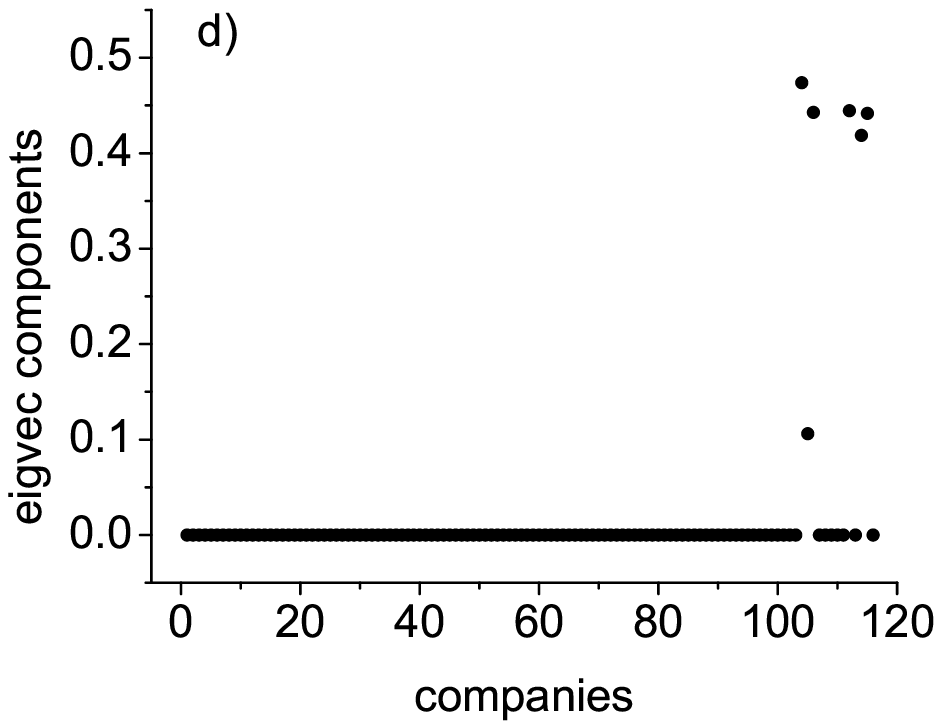}
\caption{Eigenvector components for the Energy sector a) for $p=1$, 
and b) for $p=0.024$, and for the Utilities sector c) for $p=1$, and
d) for $p=0.024$.}
\label{fig:eigs}      
\end{figure}

\begin{figure}[!h]
\centering
\includegraphics[width=265pt]{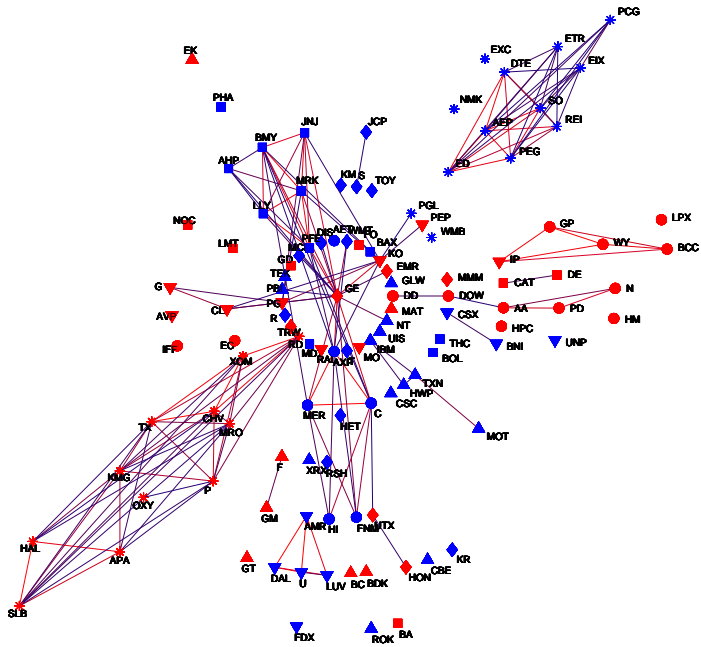}
\caption{The network for occupation $p=0.024$. The Energy sector is 
pointing to South-West and the Utilities sector to North-East direction. 
Different symbols correspond to different business sectors according to 
Forbes classification.}
\label{fig:graph}      
\end{figure}

For small $p$ the next largest eigenvalues have a clear meaning: They
correspond to highly correlated business sectors which emerge first in
the network as isolated clusters. These eigenvalues are inherited to
the $p=1$ case (though their ranking can change). Fig. \ref{fig:eigs} 
shows two such eigenvalues; for the fully connected graph ($p=1$) 
and for low occupation, $p=0.024$. This value of $p$ corresponds to 
160 links in the network at which stage its structure is shown in 
Fig. \ref{fig:graph}. We can conclude that the visually well-separated 
clusters for low occupation $p$ induces a large eigenvalue in the 
category iii) which is inherited to the final fully connected case.

\section{Subgraph Intensity and Coherence}
\label{sec:intensity}

In order to study the clustering properties in more detail, let us 
consider any cluster or subgraph $g$ in these networks by defining
two additional measures. To characterise how compact or tight the 
subgraph is, we use the concept of subgraph \emph{intensity} $I(g)$ 
introduced earlier in~\cite{intensity}. This measure allows us to 
characterise the interaction patterns within clusters. By denoting 
$v_g$ the set of nodes and $\ell_g$ the set of links in the subgraph 
with weights $w_{ij}$, we can express subgraph intensity as the 
\emph{geometric mean} of its weights as

\begin{equation} 
I(g)=\left(\prod_{(ij)\in \ell_g} w_{ij}\right) ^{1/|\ell _g|}. 
\label{eq:geom_mean} 
\end{equation} 

However, with this definition the subgraph intensity $I(g)$ may turn 
out to be low because one of the weights is very low, or all the 
weights are low. In order to distinguish between these two extremes, 
we use the concept of subgraph \emph{coherence} $Q(g)$ introduced in 
\cite{intensity} and defined as the ratio of the geometric mean to the 
arithmetic mean of the weights:

\begin{equation} 
Q(g) = I|\ell _g|/\sum_{(ij)\in \ell_g} w_{ij}. 
\label{eq:coherence} 
\end{equation} 

This coherence measure gets values in the $[0,1]$ interval and is 
close to unity only if the subgraph weights do not differ much, 
i.e. they are internally coherent. To compare the intensity and 
coherence values of various clusters, we establish a reference, 
consisting the entire market. In other words, we take all the $N$ 
nodes and $N(N-1)/2$ links making up the entire network $G$, and 
then with Eqs. \ref{eq:geom_mean} and \ref{eq:coherence}  
calculate $I(G)$ and $Q(G)$, respectively. Here we will use the 
relative quantities, i.e. \emph{relative cluster intensity} for 
cluster $g$, given by $I(g)/I(G)$, and \emph{relative cluster 
coherence}, given by $Q(g)/Q(G)$.

We will apply these measures to the same set of 116 NYSE-traded stocks 
from the S\&P 500 index and devide the stocks into clusters by using the 
same business sector labels for each stocks as above \cite{forbes}. 
Given these labels for each stock, we determine the subgraph intensity 
and coherence to gauge how stocks belonging to a given business sector 
behave as a function of time. Now let us consider a cluster $g_n$, 
constructed such that all of its nodes $v_g$ belong to the same business 
sector and $n$ denotes the number of nodes in this cluster. Then we 
add all the $n(n-1)/2$ links corresponding to the interaction strengths 
between nodes within $g_n$. In one extreme, if all the link weights 
are equal to unity, every node in $g_n$ interacts maximally with 
its $n-1$ neighbours. In the other extreme, if one or more of the 
weights are zero, the subgraph intensity for the 
\emph{fully connected subgraph} $g_n$ tends to zero because the 
original topological structure no longer exists. While this may seem
extreme, it is important to realize that the companies of any given
business sector are expected to interact, at least to some extent,
with all other companies within the sector.  In practise, however, it
rarely happens that we would have a weight $w_{ij}=0$ exactly.

\begin{figure}
\centering
\includegraphics[height=5cm]{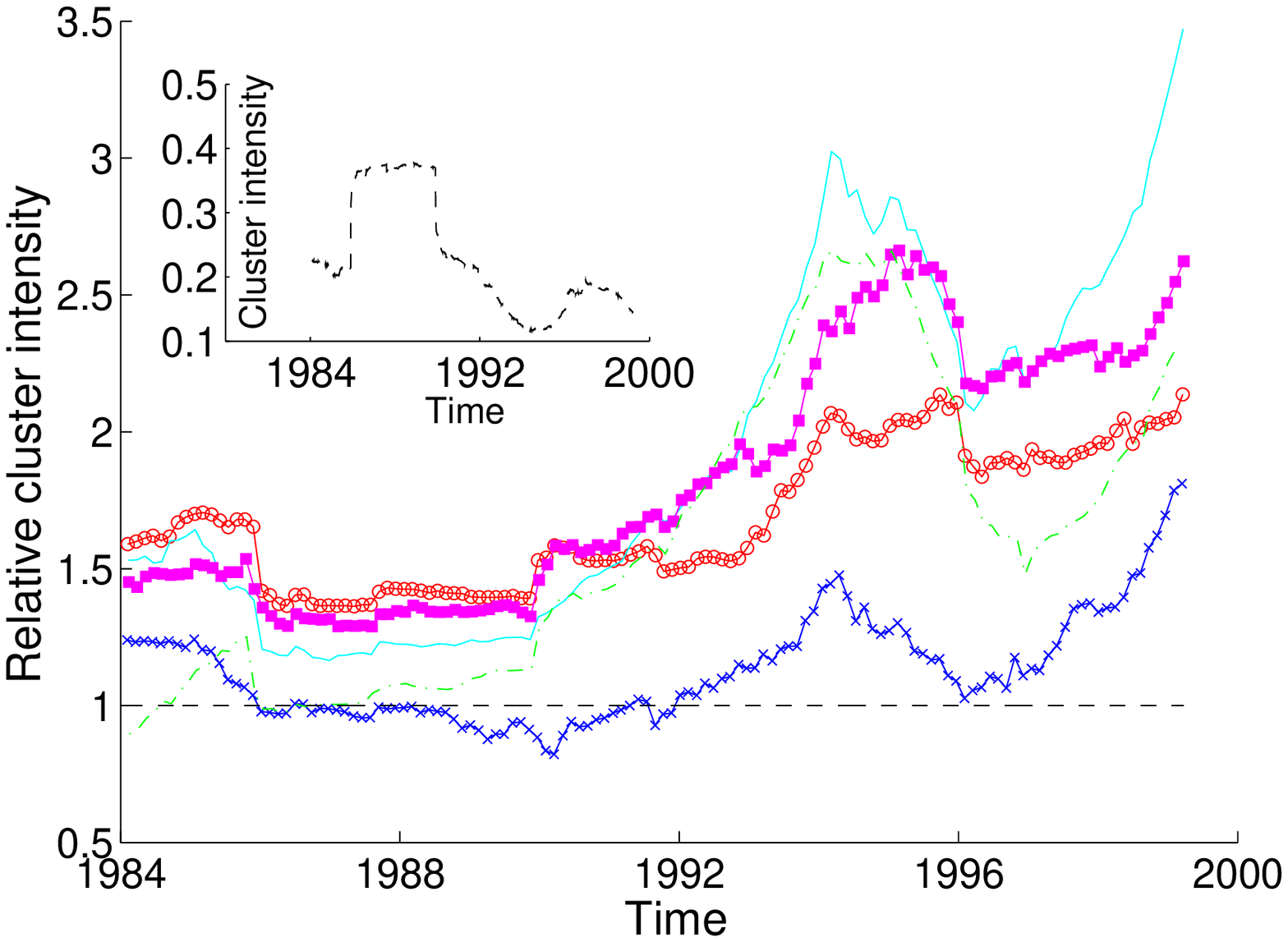}
\includegraphics[height=5cm]{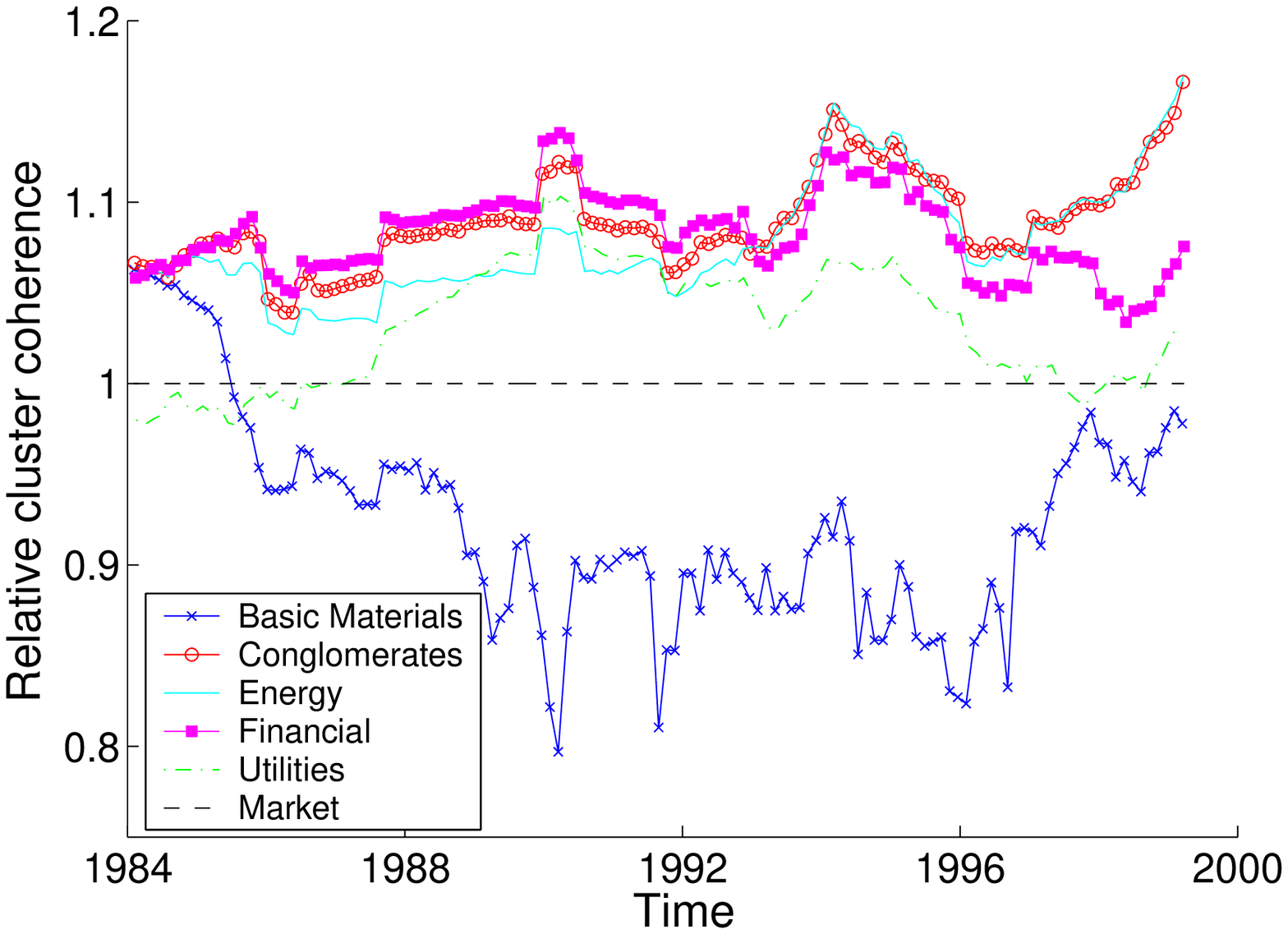}
\caption{(a) Relative cluster intensity as a function of time 
for select clusters. In the inset: The absolute cluster intensity 
for the whole market used for normalisation. (b) Relative cluster 
coherence as a function of time.}
\label{fig:int}      
\end{figure}

In Fig. \ref{fig:int}(a) we show the relative cluster intensity as a 
function of time for selected business sector clusters. Here the 
values above unity indicate that the intensity of the cluster is 
higher than that of the market. This implies that in most cases 
stocks belonging to a given business sector are tied together in 
the sense that intra-cluster interaction strengths are considerably 
stronger than the whole market interaction. In the inset of 
Fig. \ref{fig:int}(a) we have depicted the absolute cluster intensity 
for the whole market, which shows high values roughly between 1986 and 1990. 
This is caused by stock market crash (Black Monday, 1987) when the market 
behaves in a unified manner. It should be noted here that although 
the crash is a localized event, in our analysis it covers an extended 
period due to the moving window length being four years. From 
Fig. \ref{fig:int}(a) we also see that the crash compresses the 
relative cluster intensities, which means that the cluster-specific 
behaviour is temporarily suppressed by the crash. After the market 
recovers, the clusters regain their normal structural features 
\cite{Nikkei2004}.

In Fig. \ref{fig:int}(b) we show the relative coherence as a 
function of time for selected business sector clusters. All clusters 
except Basic Materials turn out to be more coherent than 
the market. One possible explanation is that for Basic Materials 
the industry classification scheme is too course, because in finer 
classification this sector includes stocks diversely from  Metal Mining, 
Paper, Gold \& Silver and Forestry \& Wood Products. 
Consequently, it is not that surprising that the cluster intensity 
remains low, at times even falling below the market reference. 
Similarly, the low coherence values indicate that there are stocks 
in this cluster with very high correlations due to those belonging 
to the same industry, such as gold mining, but also very low due 
to companies belonging to different industries. In conclusion, 
our results indicate that, in most cases, stocks belonging to the 
same business sector have higher intensity and more coherent 
intra-cluster than inter-cluster interactions. 

Support from the Academy of Finland (Center of Excellence programme), OTKA
T049238 and COST P10 is acknowledged.



\end{document}